# Photo-oxidation of Graphene in the Presence of Water

*Nobuhiko Mitoma,[†] Ryo Nouchi,*[,‡] and Katsumi Tanigaki[†]*

[†]Department of Physics and WPI-AIMR, Tohoku University, Sendai 980-8577, Japan

[‡]Nanoscience and Nanotechnology Research Center, Osaka Prefecture University, Sakai 599-8570, Japan

*Corresponding Author

E-mail: r-nouchi@21c.osakafu-u.ac.jp

Telephone/Fax number: +81 72 254 8394

**ABSTRACT**: Oxygen molecules are found to exhibit non-negligible reactivity with graphene under strong light irradiation in the presence of water. The reaction is triggered by the laser Raman spectroscopy measurement itself, and the *D* band (ca. 1340 cm$^{-1}$) becomes larger as the laser irradiation is prolonged. The electronic transport properties of the graphene derivative are also investigated and both the electron and hole mobility are found to be reduced. These results are attributed to oxidation of graphene. This primitive modification method can be exploited to manipulate the structural and electronic properties of graphene.

Graphene is a classic yet new material.[1-3] Much research has revealed the chemical and physical properties of graphene, and it is considered a potential candidate for future electronic devices due to its massless Dirac fermion properties.[4] Although graphene has a significantly high charge carrier mobility,[5,6] it is difficult to apply it to a switching device, because it is a zero bandgap semimetal. Introducing defects into the hexagonal lattice is expected to open a bandgap in graphene. The "defect" includes $sp^2$ to $sp^3$ rehybridization, which is induced by covalently bound species on graphene surfaces. Therefore, hydrogenation and other chemical modifications of graphene have been widely studied.[7-11] To investigate the amount of defects in the graphene plane, Raman scattering spectroscopy, which is known as a non-destructive characterization method for structural and electronic properties of graphene, provides much useful information.[12] The Raman peak appearing at around 1340 cm$^{-1}$, the so-called *D* band, is indicative of structural disorders or chemical modification of graphene planes. However, in line with the fact that light irradiation promotes chemical reactions of graphite and graphene,[13-16] the laser Raman spectroscopy measurement itself seems to accelerate a photochemical reaction with chemicals existing in the surrounding environment such as oxygen and water. The universally existing small *D* band in mechanically exfoliated graphene was suggested to be the result of contamination by hydrocarbons, which occasionally form C-C bonds to graphene.[9] Herein, we propose the origin of the *D* band: the photo-oxidation of graphene in the presence of water triggered by the Raman spectroscopy measurement itself. The oxidation of graphene under thermal or photochemical conditions have been reported,[17-19] and the former reaction was found to occur preferentially at point defects. The high reactivity in graphene defects or edges is explained by the existence of unpaired electrons. However, oxidation of graphene in the presence of water is not restricted to the edge region as is the case for the UV/ozone oxidation,[18] and oxidation can occur even in the non-edge region of graphene, as shown in this paper. This could implicate Raman spectroscopy, a widely used structural evaluation method, as a destructive technique in some cases.

Single layer graphene (SLG) flakes were prepared on a heavily doped silicon substrate with a 300-nm-thick thermal oxide layer by micromechanical cleavage.[3] Both "dry" and "wet" $SiO_2$/Si substrates were used to support the graphene flakes. The dry substrate was prepared by annealing as-prepared $SiO_2$/Si substrates at 110 °C under ambient conditions for 2 h to eliminate residual water molecules on the substrate. While this process is not sufficient to completely remove residual water molecules on $SiO_2$ surfaces, the resulting substrate is definitely drier than "wet" substrates. The wet substrate was prepared as follows. The $SiO_2$/Si substrate was first treated with a semiconductor cleaning liquid, followed by exposure to oxygen plasma to remove hydrocarbon contaminants on the surface. Such treatment is known to increase the hydrophilic properties of these substrates.[20] After the plasma treatment, the substrate was immersed in deionized water for 1 min and was dried by blowing air. SLG flakes were then immediately formed on the substrate in ambient air. Hereafter, we refer to these samples as single layer graphene on a dry substrate (SLGD) and single layer graphene on a wet substrate (SLGW). The number of graphene layers was confirmed by its optical contrast and Raman spectra.[12] Raman spectroscopy measurements were conducted under ambient conditions. The data were collected every 10 min and the laser light irradiation (green laser with an excitation energy of 2.3 eV; output power of 7.7 mW; focused beam diameter of 5 μm) was prolonged up to more than 300 min. No difference was detected by optical microscopy of the graphene after laser irradiation.

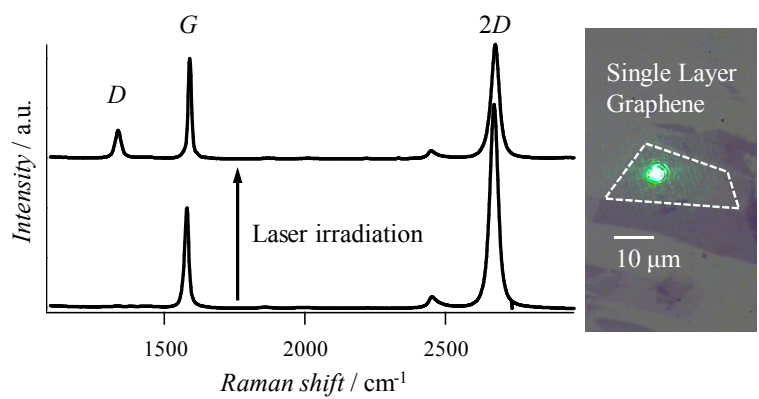

**Figure 1.** Raman spectra normalized to the peak intensity of the *G* band before and after laser irradiation for 5 h. A SLGW was used for the experiment. The spectra are shifted for clarity. The inset shows an optical microscope image of the sample. The bright green spot appearing on the graphene plane is the laser spot.

Figure 1 shows Raman spectra from SLGW. The spectra are normalized to the peak intensity of the *G* band. All the Raman measurements were carried out in ambient air. The excitation laser was irradiated on the basal plane of a graphene sample to minimize the influence of graphene edges (see the inset in Figure 1) for up to 5 h. The *D* band intensity was found to increase as the laser irradiation was prolonged. Wavenumber shifts of the *G* (1581 to 1591 cm$^{-1}$) and 2*D* (2671 to 2676 cm$^{-1}$) bands were also discernible. These shifts are signs of charge carrier doping to graphene.[21,22] The 2*D*/*G* intensity ratio decreased as the laser irradiation was prolonged, which is also a signature of charge carrier doping (Figure S3 in Supporting Information). Here, chemisorption and physisorption of oxygen may occur simultaneously. These two events can be distinguished by Raman spectra; chemisorbed species induce a clear *D* band, while physisorbed species do not.[15] Raman spectra obtained from bi-layer graphene (BLG) on a wet substrate were also investigated. A clear difference compared to those from SLGW is that the *D* band did not appear even after laser irradiation for 4 h (Figure S4 in Supporting Information). This difference in chemical reactivity between SLG and BLG is explained as follows. The surface ripples of SLG are considered to reduce the activation energy for the formation of $sp^3$ bonds on a surface. BLG has fewer ripples compared to SLG, which is attributable to the less flexibility due to its double thickness.[23,24]

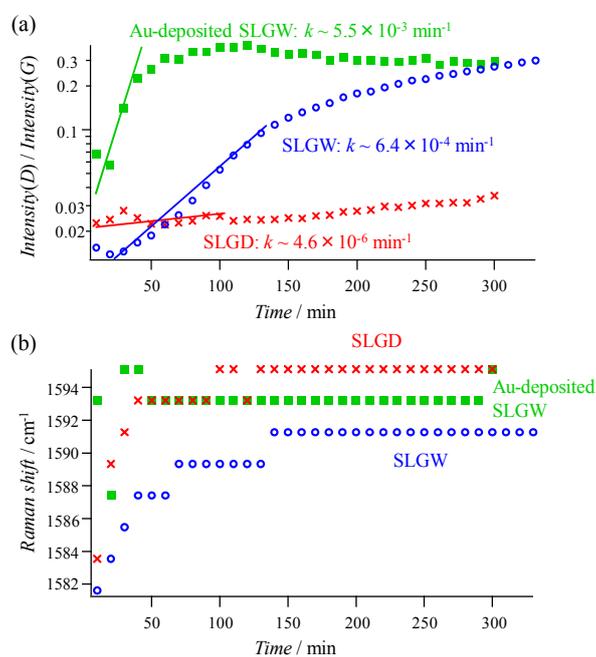

**Figure 2.** (a) Time evolution of the *D*/*G* intensity ratio from three types of graphene flakes. Data points represent experimentally obtained values, and solid lines are exponential fits to the experimental values. (b) Time evolution of the *G* peak shift obtained from three samples. The circles, crosses, and squares represent the SLGW, SLGD, and Au-deposited SLGW samples, respectively.

Figure 2(a) is the time evolution of the $D/G$ intensity ratio obtained from three types of SLG samples. The $D/G$ intensity ratio of the SLGW was clearly enhanced as the laser irradiation was prolonged, while that of the SLGD was less than 0.05 even after laser irradiation for 5 h. These results indicate that water molecules are essential to induce the $D$ band. It should be noted that a laser of weak output power, e.g., less than 1 mW, is used for conventional Raman spectroscopy measurements; however, it can be said that laser irradiation with output power of 7.7 mW induced sufficiently few structural disorders in this study as seen in the SLGD results. SLGW, on which a 0.5-nm-thick discontinuous Au film was deposited, had an even faster enhancement of the $D$ band intensity. Discontinuity of the film was confirmed both from the electron transport properties and atomic force microscopy images (Figures S5 and S6 in Supporting Information). This faster enhancement might be caused by an evanescent field generated underneath the Au particles. This electromagnetic field strongly enhances the laser intensity, i.e., photon flux $f$, leading to the $D$ band intensity enhancements. A sample of Au-deposited SLGD was also measured; however, no obvious $D$ band enlargement was evident (Figure S2 in Supporting Information). This result also suggests that water molecules are essential for the $D$ band induction.

Figure 2(b) shows the $G$ peak position which indicates the charge carrier doping. The clear shift of the SLGW and SLGD is considered to be caused by physisorption of oxygen molecules from ambient air triggered by the laser irradiation.[25] The Raman peak position shift of Au-deposited SLGW was small compared to those of the SLGW and SLGD, and the $2D/G$ ratio was almost constant during light irradiation. These results may be ascribable to charge density pinning caused by the deposited Au particles.[26-28] The charge density pinning and the evanescent field generation should occur at the same position; thus, the Raman peak shift can be very small in the Au-deposited SLGW sample.

The time evolutions of the photochemical reaction can be divided into two stages: an abrupt increase in the $D/G$ intensity ratio, followed by a gradual decrease. The abrupt increase of the $D/G$ intensity in the former part can be understood as follows. The two prominent $D$ and $G$ bands originate from the breathing mode of K point phonons of $A_1'$ symmetry and the $E_{2g}$ phonon of $sp^2$ C atoms, respectively. The $D/G$ intensity ratio is proportional to the reciprocal of the average crystallite size of graphene with $sp^2$ C networks.[29] A chemically modified site breaks the local π conjugation, and thus the time evolution of the intensity ratio is expected as

$$I_D(t)/I_G(t) \propto \exp[kt]-1, \quad (1)$$

$$k = n_W \sigma f, \quad (2)$$

where $n_W$ is the number of water molecules, $\sigma$ is the photoreaction cross section, and $f$ is the incident photon flux (see Supporting Information). Such an exponential dependence of photoreaction was also observed in alkali metal adsorbed graphite systems.[13] Solid lines in Figure 2(a) are least square fits to the experimental data using the above equations. The slope represents the probability of the photochemical reaction in any time interval $\Delta t$. The reaction is dominated by the number of water molecules which assist oxidation as discussed in the next paragraph. The absence of the $D$ band seen in SLGD implies a weak interaction between graphene and oxygen molecules.[25] The Raman laser itself partially creates a photostationary state in which optically excited graphene transfers hot electrons to physisorbed oxygen molecules on the graphene surface, which results in the hole doping. At the same time, the high power laser heats up the sample and photothermal desorption of water molecules might occur. This may be observed as the reduction in the reaction probability in the latter part. The possibility of the *D* band decrease due to desorption of covalently bound oxygen[30] is excluded here, because a thermal treatment in Ar-$H_2$ atmosphere at 1000 °C was reported to be necessary to reduce graphene oxide.[31] The temperature of the sample is estimated to be around 100 °C at most considering the 7.7 mW of laser power that was focused on a 5-μm diameter spot.[32]

The aqueous oxygen redox couple, $O_2 + 4H^+ + 4e^- \leftrightarrow 2H_2O$, is likely to be formed on the surface of graphene.[33] Oxygen molecular anions formed by laser-induced hot electron transfer from graphene are stabilized by water solvation and electrostatic binding to the $SiO_2$ surface.[25] The oxygen anions stabilized in the presence of water can be used to oxidize graphene. The plasma treatment employed for the preparation of the wet substrates is known to change charge inhomogeneity and to increase roughness of the substrate surface. The possibility that such effects reduce the activation energy for oxidation of graphene still remains; further study necessary to clarify these effects on the graphene oxidation. A possibility of hydrogenation can be excluded here because heat treatment at 200 °C, at which hydrogen atoms are desorbed to some extent,[32] was found not to reduce the *D* band intensity. Water is transparent to visible light, and thus it cannot be decomposed by single photon absorption from the green laser at 2.3 eV used in the present study. The possibility of water decomposition by multiphoton absorption is also considered to be limited with the photon energy of 2.3 eV.[34] Together with the fact that a control experiment where a SLGW sample was sealed in a small chamber filled with Ar gas showed no obvious *D* band induction, these considerations support that the reaction discussed in the present study is oxidation assisted by water molecules.

The electron transport properties of graphene field effect transistors (FETs) were also investigated. The FETs used in this study were fabricated by a conventional liftoff technique. Graphene flakes were deposited on the dry substrate to diminish the effect of residual water between the graphene sheet and the substrate surface.[35,36] An electron beam resist was spin-coated on the substrate and annealed at 180 °C for 3 min on a hot plate. An electron beam lithography system was used to pattern electrodes. After developing the resist, Cr (1 nm thick as an adhesion layer) and Au (30 nm thick) layers were fabricated by vacuum deposition.

All the FET measurements were carried out at room temperature in high vacuum (ca. $10^{-2}$ Pa). The channel length and width of fabricated graphene FETs were 10 and 2 μm, respectively. All the FETs were annealed at 110 °C for 2 h in ambient air to desorb water molecules on the graphene basal plane prior to electric measurements. After the first FET measurements, the FETs were immersed in water under dark conditions at room temperature for 48 h, then the annealing procedure at 110 °C was repeated, followed by the second FET measurement. No obvious evidence of chemical reaction between graphene and water molecules was found in this case. A large hysteresis in transfer characteristics is generally induced by residual water molecules.[36] Since water molecules could be desorbed under high vacuum condition, such a hysteretic behavior was not observed in the present devices. Chemisorption must result in a decrease in charge carrier mobility, because π electrons on the graphene planes are used to form $sp^3$ covalent bonds.[8] The mobility of the device increased after the immersion in the dark condition due to the removal of contaminants on graphene plane (Figure S8 in Supporting Information).

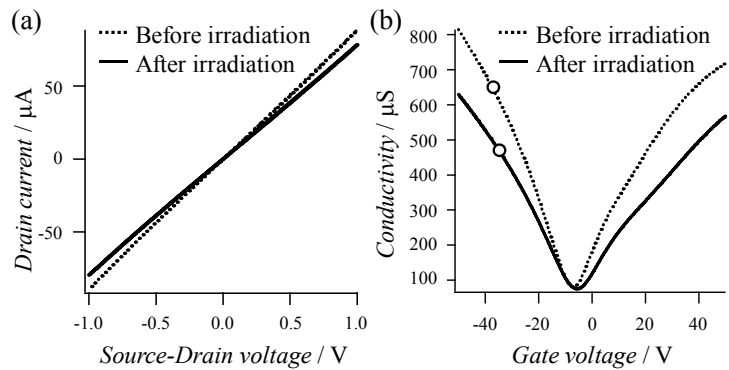

**Figure 3.** (a) Output and (b) transfer curves obtained before and after the Xe lamp irradiation in water. The output curves were taken at the minimum-conductivity point [$V_G$ = -7 V (dashed curve), $V_G$ = -5 V (solid curve)]. The channel length and width of the graphene FET are 10 and 2 μm, respectively. Circles indicate the positions at which hole mobility was extracted.

Next, the same FET was irradiated with light in water for 33 h using a Xe lamp in which a small amount of Hg vapor was sealed (light power of 60 nW μm$^{-2}$). The Xe lamp contains UV light whose energy is higher than 6.5 eV and thus can decompose water molecules,[37] i.e., a certain amount of hydrogen and oxygen molecules are generated in water. Figure 3 shows the electron transport properties of the FET measured using the two-terminal current-voltage method before and after the light irradiation. Obvious changes in electron transport properties were observed in all the fabricated FETs. The field-effect mobility of the device was extracted from the dashed curve in Figure 3(b) to be around 9560 cm$^2$ V$^{-1}$ s$^{-1}$ in hole-conduction regions. After irradiation with the Xe lamp, the mobility was decreased to ca. 7510 cm$^2$ V$^{-1}$ s$^{-1}$, and the resistivity at the minimum-conductivity point increased slightly. These degradations imply the presence of covalently bound species or structural disorders on the graphene plane.[38,39] In the present case, oxidation should be induced by the UV irradiation.[18]

In summary, enlargement of the *D*/*G* intensity ratio of graphene on a water rich substrate was observed during laser Raman spectroscopy measurements. This effect is significantly enhanced in the presence of metal particles on graphene. These results are explained in terms of the photo-oxidation of graphene in the presence of water. It should be noted that Raman spectroscopy is a widely used structural evaluation method; however, this technique can induce a structural change of graphene. The photo-oxidation can be mimicked by irradiating UV light on graphene immersed in water, and the obtained graphene derivative exhibits a reduction in electron and hole mobility. The experimental method presented here is easily applicable to a variety of graphene FET systems, and would be a useful contribution to the electronics with graphene.

**Notes**

The authors declare no competing financial interest.

**Acknowledgments**

This work was partially supported by the Weaving Science Web beyond Particle-Matter Hierarchy, Global Centers of Excellence program, Tohoku University; the World Premier International Research Center Initiative; and Special Coordination Funds for Promoting Science and Technology from the Ministry of Education, Culture, Sports, Science and Technology, Japan. The authors express their deep gratitude to the Department of Instrumental Analysis, School of

ABBREVIATIONS

SLGD, single layer graphene on a dry substrate; SLGW, single layer graphene on a wet substrate; FET, field effect transistor.

# Supporting Information

# Photo-oxidation of Graphene in the Presence of Water

Nobuhiko Mitoma,[†] Ryo Nouchi,*[,‡] Katsumi Tanigaki[†]

[†]Department of Physics and WPI-AIMR, Tohoku University, Sendai 980-8577, Japan

[‡]Nanoscience and Nanotechnology Research Center, Osaka Prefecture University, Sakai 591-8570, Japan

*To whom correspondence should be addressed: r-nouchi@21c.osakafu-u.ac.jp

**Sample fabrication**

Crystals of Kish graphite were purchased from Covalent Materials Corporation. Single- and bi-layer graphene (SLG and BLG) flakes were prepared on a heavily doped silicon substrate with a 300-nm-thick thermal oxide layer (Kojundo Chemical Laboratory Co., Ltd.) using the crystal and Scotch® tape.[1] Both "dry" and "wet" substrates were used in this study. The dry substrate was prepared by annealing as-prepared $SiO_2$/Si substrates at 110 °C under ambient conditions for 2 h to eliminate residual water molecules on the substrate. The wet substrate was prepared as follows. The $SiO_2$/Si substrate was first treated with a semiconductor cleaning liquid (Semicoclean23, Furuuchi Chemical Co., Ltd.), followed by exposure to oxygen RF plasma with an output power of 40 W for 3 min to remove hydrocarbon contaminants on the surface. Such treatment is known to increase the hydrophilic properties of these substrates (see Reference 2 and Figure S1). After the plasma treatment, the substrate was immediately immersed in deionized water for 1 min and was dried by blowing air; SLG and BLG flakes were then immediately formed on the substrate. The number of layers of graphene flakes was confirmed by the optical contrast and the shape of the Raman spectrum.[3] Hereafter, we refer to these samples as single-/bi- layer graphene on a dry substrate (SLGD/BLGD) and on a wet substrate (SLGW/BLGW).

Field effect transistors were fabricated by a conventional lift-off technique. Graphene flakes were deposited on the dry substrate to exclude the effect of residual water between the graphene sheet and the substrate surface. An electron beam resist (ZEP520A, Nippon Zeon Co., Ltd.) was spin-coated on the substrate and annealed at 180 °C for 3 min on a hot plate. An electron beam lithography system (ELS7500S, Elionix Inc.) was used to pattern electrodes. After developing the resist, Cr (1 nm thick as an adhesion layer) and Au (30 nm thick) layers were deposited.

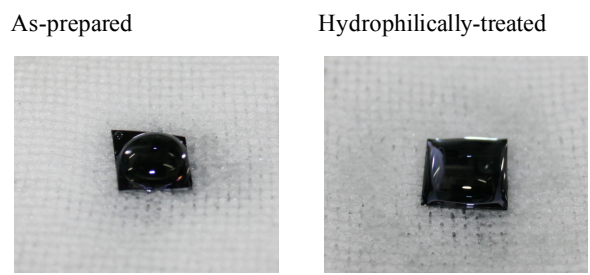

**Figure S1.** Hydrophilic properties of as-prepared and hydrophilically treated substrates. The contact angle of water was clearly reduced after hydrophilic treatment.

**Raman spectra and influence of water**

Figure S2 shows the Raman spectra obtained from four different samples. These results clearly show that the presence of water molecules is essential to induce the *D* band. Figure S3 presents the 2*D*/*G* intensity ratio of the samples mentioned in the main text (see also Figure 2). The 2*D*/*G* intensity ratio is used as an index of charge carrier doping.[4]

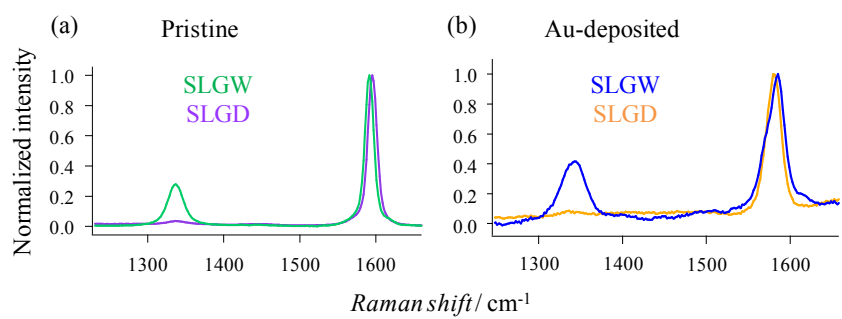

**Figure S2.** Comparison of Raman scattering spectra for graphene deposited on wet and dry substrates. The results indicate that water molecules are essential to induce the *D* band. Laser irradiation was maintained for 5 h for pristine graphene and 30 min for Au-deposited graphene.

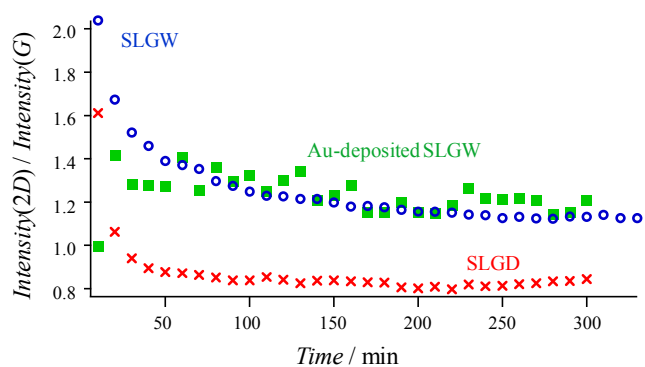

**Figure S3.** Time evolution of the 2*D*/*G* intensity ratio obtained from SLGW, SLGD and Au-deposited SLGW samples.

**Raman spectra of SLGW and BLGW**

The difference in chemical reactivity between SLG and BLG can be explained in terms of the ripples and flexibility of the basal plane. SLG has more ripples and is thus more flexible than BLG, which reduces the activation energy for the formation of $sp^3$ bonds on the graphene surface.[5,6] This is directly reflected in the time evolution of its Raman spectra as shown in Figure S4.

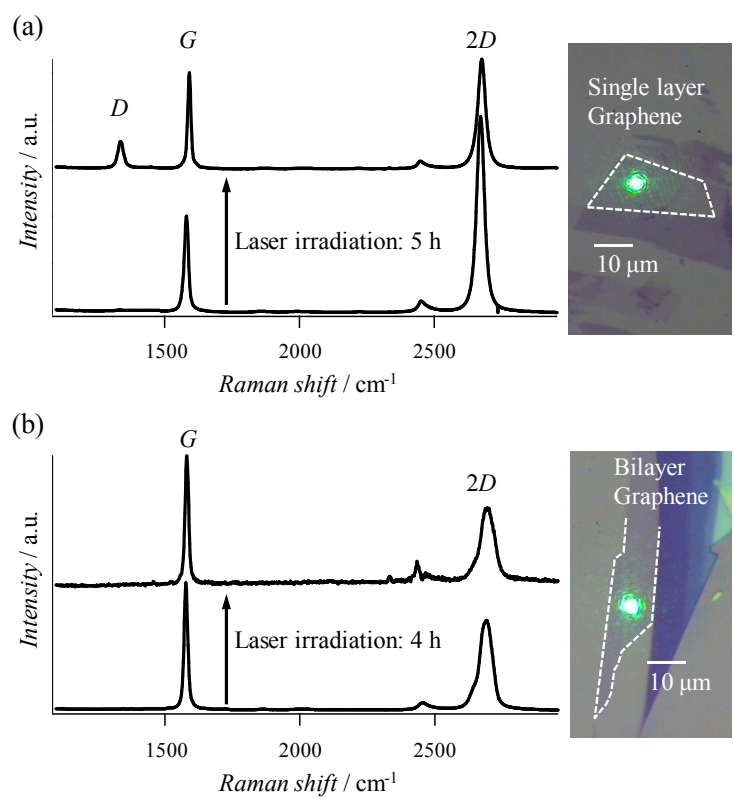

**Figure S4.** Raman spectra normalized to the peak intensity of the *G* band before and after laser irradiation. (a) SLGW and (b) BLGW were used for the experiment. The spectra are shifted for clarity. The insets show optical microscope images of the samples. The bright spot appearing on the graphene plane is the laser spot.

**Au nanoparticles deposited on graphene**

A discontinuity of the deposited Au film on graphene was confirmed as follows. If a continuous Au film is formed on graphene, then the electric current mainly flows through the Au percolation path, and the FET would not exhibit gate voltage dependence. However, the transfer curve shown in Figure S5 exhibits a gate voltage dependence, although it has low conductivity due to charge carrier scattering from the Au particles.[7] Atomic force microscopy also confirmed the discontinuity of the Au film (Figure S6).

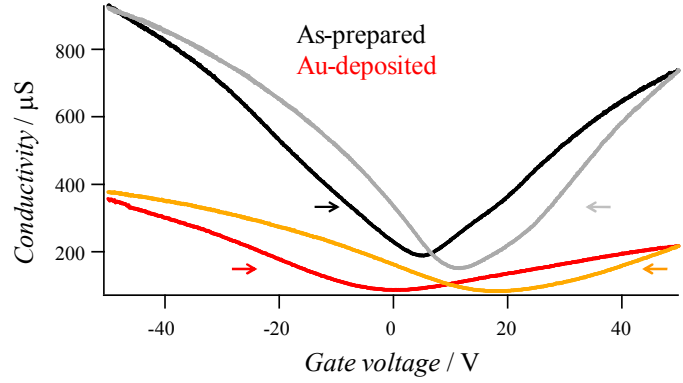

**Figure S5.** Transfer characteristics of a graphene FET before and after Au deposition. The gate-voltage dependence persists after Au deposition, which indicates that a significant percolation path of the Au film was not formed.

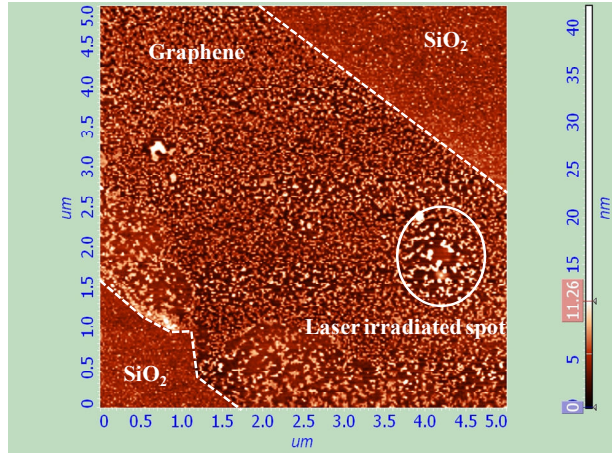

**Figure S6.** Atomic force microscope image of the 1-nm-thick Au layer deposited on a graphene sample. The circle indicates the laser-irradiated region. Coalescence of Au particles is evident, which could be due to the increase in the substrate temperature during laser irradiation.

**Time evolution of the *D* band intensity**

Here we propose a simple two-state model. We assume that the intensity of the *D* band is proportional to the number of carbon atoms chemically modified in the presence of water molecules. Supposing that the probability of photochemical reaction in any interval $\Delta t$ is $n_W \sigma f \Delta t$, the following equations result:

$$N = n(0) = n(t) + n_D(t), \tag{S1}$$

$$n(t + \Delta t) = n(t) - n(t) n_W \sigma f \Delta t, \tag{S2}$$

where $N$ is the number of C atoms in the laser-irradiated region, $n(t)$ is the number of $sp^2$ C atoms and $n_D(t)$ is the number of $sp^3$ C atoms. $n_W$, $\sigma$ and $f$ represent the number of water molecules, reaction cross section and incident photon flux, respectively. From these equations, $n(t)$ can be expressed as

$$n(t) = n(0)(1 - n_W \sigma f \Delta t)^{\frac{t}{\Delta t}}$$
$$\rightarrow N \exp[-n_W \sigma f t] \qquad (\Delta t \rightarrow 0) \quad . \tag{S3}$$

From Equations (S1) and (S3),

$$n_D(t) = N(1 - \exp[-n_W \sigma f t]) \quad . \tag{S4}$$

The $D$ and $G$ band intensities are considered to be proportional to $n(t)$ and $n_D(t)$, respectively. Thus the time evolution of the $D/G$ intensity ratio is expected as

$$I_D(t)/I_G(t) \propto n_D(t)/n(t)$$
$$= \exp[kt] - 1, \tag{1}$$

$$k = n_W \sigma f \quad . \tag{2}$$

**Carrier transport properties of the FET**

Electronic transport measurements were conducted using a parameter device analyzer (B1500A, Agilent Technologies). All measurements were carried out at room temperature. The transport properties of the devices were significantly affected by the level of vacuum, which determines the number of water molecules adsorbed on graphene, as shown by the shape of the transfer curves in Figure S7. Figure S8 indicates that the water-induced hysteresis[8] was almost removed in high vacuum. To remove the effect of water, substrates were annealed at 110 °C under ambient conditions for 2 h, and the measurements were performed under high vacuum conditions.

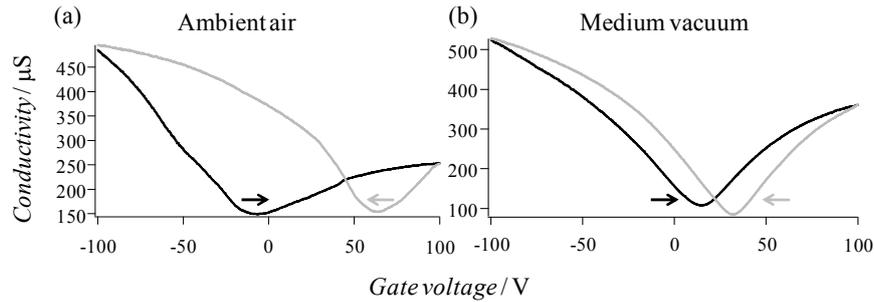

**Figure S7.** Hysteretic behavior obtained from SLGD sample immersed in water. The transfer curves were measured in (a) ambient air and (b) 10 Pa, respectively. (a) and (b) were obtained from the same device. Arrows represent the voltage sweep direction.

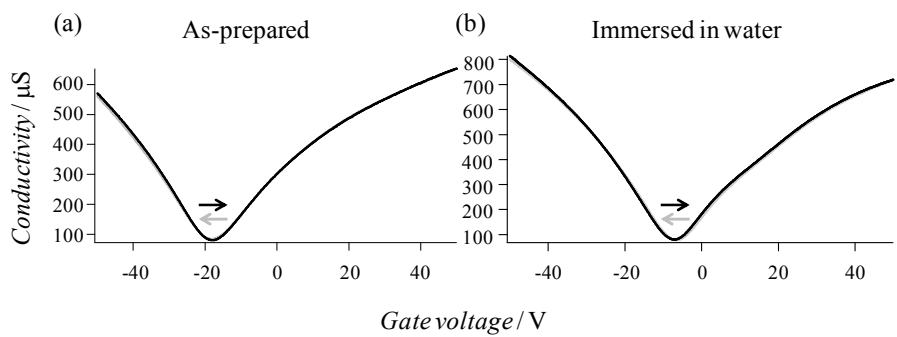

**Figure S8.** Transfer characteristics obtained from the FET device in the main text (see Figure 3 in the main text). These measurements were performed for the (a) as-prepared and (b) immersed in water under dark conditions samples. To remove the effect of adsorbed water on graphene, the measurements were carried out under high vacuum (ca. $10^{-2}$ Pa). The charge carrier mobilities measured at each step are given in Table S1. A positive shift of the minimum-conductivity point and enhancement of the mobility were observed. These changes may be due to the removal of the electron beam resist residue on the graphene sheet.

**Table S1.** Charge carrier mobilities before and after the immersion in water under dark conditions. The obtained values are field-effect mobilities, and include the contribution of electrode contact resistances.

|  | As prepared | Immersed in water |
|---|---|---|
| Electron mobility ($cm^2\ V^{-1}\ s^{-1}$) | 8095 | 9559 |
| Hole mobility ($cm^2\ V^{-1}\ s^{-1}$) | 6299 | 7406 |

**Acknowledgment**

The authors thank Y. Nishikawa of Quantum Design Japan for obtaining the atomic force microscope image.